\documentclass[fleqn,usenatbib]{mnras}

\usepackage{newtxtext,newtxmath}
\usepackage{bm}
\usepackage{amsmath}
\usepackage{multicol}

\usepackage[T1]{fontenc}

\DeclareRobustCommand{\VAN}[3]{#2}
\let\VANthebibliography\thebibliography
\def\thebibliography{\DeclareRobustCommand{\VAN}[3]{##3}\VANthebibliography}

\usepackage{graphicx}	% Including figure files
\usepackage{amsmath}	% Advanced maths commands
\title[Mass-Ratio Distribution II]{The Frequency and Mass-Ratio Distribution of Binaries in Clusters II: radial segregation in the nearby dissolving open clusters Hyades and Praesepe. }

\author[M. D. Albrow]{Michael D. Albrow\thanks{E-mail:Michael.Albrow@canterbury.ac.nz}\\
\\
$^{1}$School of Physical and Chemical Sciences, University of Canterbury, Private Bag 4800, Christchurch, New Zealand\\
}

\date{Accepted XXX. Received YYY; in original form ZZZ}

\pubyear{2024}

\hypersetup{draft}

\begin{document}
\label{firstpage}
\pagerange{\pageref{firstpage}--\pageref{lastpage}}
\maketitle

\begin{abstract}
We have determined the mass functions, mass-ratio distribution functions and fractions of binary stars with mass ratios above particular thresholds for radially-separated populations of stars in the nearby open clusters Hyades and Praesepe. Radial mass segregation is detected, with the populations of stars within the tidal radii having much flatter mass functions than those outside the tidal radii.
Within the tidal radii, the frequency of binary stars with mass ratio $q > 0.5$ is  50 - 75 per cent higher  for Hyades and 5 - 30 per cent higher for Praesepe.  
We also, for the first time, detect mass-ratio radial segregation.
Of the binaries for which $q > 0.5$, $\sim$80 per cent of the inner Hyades population also have $q > 0.75$, while for the extra-tidal population, the ratio is $\sim$50 per cent.
For Praesepe,  $\sim$67 per cent of the inner sample have $q > 0.75$, and $35 - 45$ per cent of the outer sample.

\end{abstract}


%%%%%%%%%%%%%%%%%%%%%%%%%%%%%%%%%%%%%%%%%%%%%%%%%%

%%%%%%%%%%%%%%%%% BODY OF PAPER %%%%%%%%%%%%%%%%%%

\section{Introduction}

Open star clusters are important targets for studying how stellar populations form and disperse in the Galaxy. Most field stars
are thought to form in clusters or looser associations, and we ought to be able to reconcile the properties of open
cluster stellar populations with those of the Galactic disk.

Properties of binary star systems offer powerful constraints on star formation and early cluster evolution \citep{Kraus2012}.
Binary stars in dense clusters provide a heat reservoir, slowing or preventing runaway core collapse through close gravitational encounters 
with single stars that (after possible exchange of partners) effectively provide kinetic energy to the single star through an increase in negative binding energy (hardening) 
of the binary \citep{Elson1987, Hut1992, Hurley2005}. Binaries can also form in cluster cores through 2-body capture interactions.  In the less-dense environments
of open clusters, these effects will be less significant, but may still play an important part in controlling the 
mass distribution of ejected (evaporated) stars. In turn, we would like to know
whether binaries of different mass ratios are subject to different
rates of disruption and capture-formation, and more-generally how the dynamical processing in cluster cores affects the
mass-ratio distribution of binaries.

Binary stars have been observed  to be more centrally concentrated than single stars in globular \citep{Albrow2001} and open clusters \citep{Childs2023}, and generally this is in accord with expectations from N-body simulations 
\citep{Geller2013}.

Previous observational studies of individual clusters have found binary mass-ratio distribution functions  that are flat \citep{Torres2021} or gently rising \citep{Reggiani2013} or falling \citep{Patience2002}. 
The recent large open-cluster study of \citet{Cordoni2023} found that while the mean behaviour of the mass ratio distribution for $q > 0.6$ is fairly flat, individual
clusters can display distribution functions that either rise or fall.
In contrast, \citet{Malofeeva2023}, finds a strongly decreasing distribution with $q$ for Pleiades, Alpha Per, Praesepe, and NGC 1039.
We note that different authors have used quite different methods for these analyses.
Theoretical models may not reproduce the observed distributions -- e.g. the N-body model of \citet{Geller2013}, while successful in reproducing
many properties of NGC 188, fails to explain the observed $q$-distribution for main-sequence stars.

There is some discrepancy over the binary mass-ratio distribution for field stars. From radial velocity studies, 
\citet{Duquennoy1991} found a distribution that peaks around q = 0.25 and declines towards higher mass ratios, while \citet{Fisher2005} and \citet{Raghavan2010} find a flat distribution with a sharp upturn towards higher mass ratios, $q \gtrsim 0.8$. \citet{El-Badry2019} confirmed this high-$q$ peak, but showed that it only exists for closely-separated binaries, $s \lesssim 600$ au.

\citet{Bate2009} and \citet{Moeckel2010} provide simulations of an open cluster that results from the collapse of a 500~$M_\odot$ molecular cloud. The resulting cluster contains 1253 stars (191 $M_\odot$ in stars)  that initially (0.3~Myr) have a binary mass-ratio distribution that rises slowly towards $q = 1$. After 10~Myr of dynamical processing, the distribution for binaries with primary mass > 0.5~$M_\odot$
has flattened a little, but remains unchanged for systems with lower-mass primaries.
 
In contrast, recent simulations by \citet{Guszejnov2023} of cluster formation from a $2 \times 10^4 \, M_\odot$ primordial cloud produce
a binary mass ratio distribution that, after $\sim$ 10 Myr,  is strongly peaked at $q \approx 0.2$, consistent with random sampling from the IMF.
The distribution is almost flat for $q >0.4$, but has a deficit of equal-mass binaries.

In this paper, we investigate the  nearby coeval open clusters, Hyades (Mel 25, Col 50) and Praesepe (M 44, Mel 88, NGC 2632).
At a mean distance of 47 pc, Hyades  is the closest open star cluster to Earth. It has a core radius of 2.7 pc \citep{Perryman1998}, mapping to an angular size of $\sim 4$ deg, and a half-mass radius of 
5.75~pc, \citet{WynEvans2022}). It also contains member stars that are dispersed across some 90 deg of the sky \citep{Meingast2019,Jerabkova2021}. \citet{Oh2020} have found that Hyades is rapidly disintegrating, estimating a current mass loss rate of at least $0.26 \, {\rm M_\odot \, Myr^{-1}}$, 
with a further 30~Myr until final dissolution. The estimated initial mass of the cluster $\gtrsim 750\, {\rm M_\odot}$. 

Praesepe, located at a distance of 185 pc, is a little more tightly bound than Hyades, with a core radius of 1.6 pc  \citep{Gao2019} which projects to 0.5 deg, and a half-mass radius of 
4.8~pc \citep{Roser2019}. It also is associated with vast tidal tails that stretch across tens of degrees on the sky 
\citep{Roser2019}.

In Paper 1 of this series \citep{Albrow2022}, we introduced a new probablistic method to measure the binary frequency and mass-ratio distribution function
for stellar populations. In this paper, we apply the method to radially segregated samples of stars from Hyades and Praesepe. 

\begin{figure}
    \includegraphics[width=\linewidth]{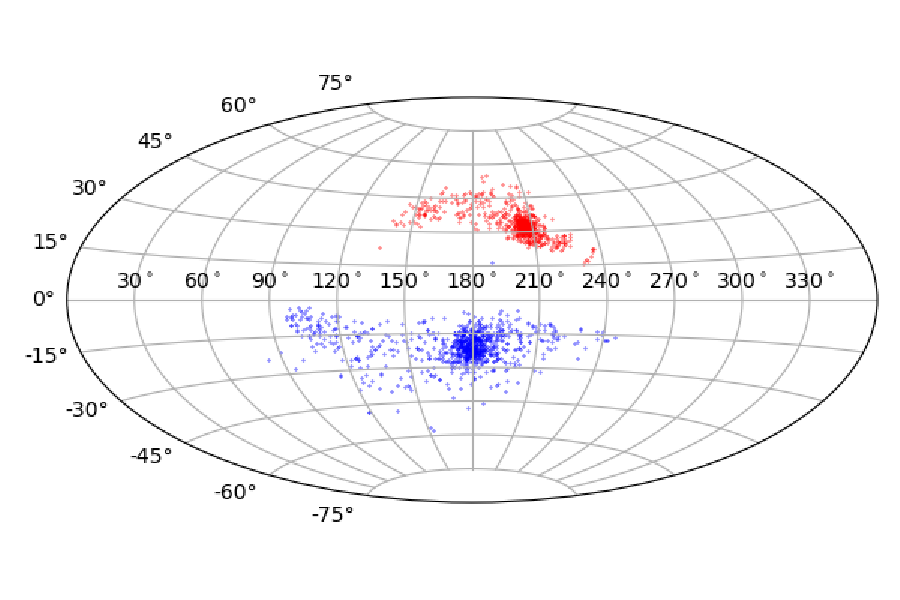}
\caption{Aitoff projection, in galactic coordinates, of the Hyades (blue) and Praesepe (red) cluster members.}
\label{fig:projection}
\end{figure}

\section{Data}\label{sec:Data}

All of the photometric data used in this paper comes from Gaia data release 3 (DR3 or EDR3) \citep{Riello2021}.  
For  Hyades, we adopted the selection of stars from the Gaia Catalog of Nearby Stars (GCNS) 
\citep{Gaia2021}, a compilation of Gaia sources within 100 pc from the Sun. Stars belonging to Hyades
were selected in the GCNS based on their distances, proper motions and radial velocities, resulting in 920 member stars.

For Praesepe, we adopt the selection of \citet{Roser2019} from Gaia DR2 (1394 stars). These were cross-matched to DR3 using the table gaiadr3.dr2.neighbourhood provided by Gaia.

To compute distances, and hence absolute magnitudes we use the simple $d(\rm pc) =  1000/\varpi({\rm mas})$ relation without priors.
The median distances of the cluster core members are 47.4 pc (Hyades) and 184.8 pc (Praesepe). 
The distribution of cluster member stars across the sky is shown in Fig.~\ref{fig:projection}.

\section{Age and Metallicity}

The age of these clusters has been determined from CMD fitting by various authors. These estimates tend to depend on the adopted metallicity
and particularly the adopted stellar rotation velocity and degree of convective core overshoot, with a higher rotation or overshoot allowing stars to turn off from the main sequence at a 
higher luminosity for a given age. 

For Hyades, estimated ages include $710$ Myr \citep{Brandner2023},  $650$  Myr \citep{Lebreton2001},  $800$ Myr \citep{Brandt2015}, and 600 - 800 Myr \citep{Gossage2018}. \citet{Lodieu2020}  have determined an age of $650 \pm 70$ Myr spectroscopically, from lithium depletion in L dwarfs.
Metallicity estimates are in the range  [Fe/H] $\approx$ 0.0 - 0.25 \citep{Brandner2023}.

Praesepe is believed to be co-eval with the Hyades and share the same metallicity \citep{Brandt2015,Gossage2018}.

For our modelling (described below), we adopt isochrones from the MESA Isochrones and Stellar Tracks project\footnote{http://waps.cfa.harvard.edu/MIST/}
\citep{Dotter2016,Choi2016,Paxton2011,Paxton2013,Paxton2015}. From our own fitting of these isochrones to the main sequence and turnoff regions of the CMD's, we
have adopted an age of 630 Myr and metallicity of  [Fe/H] = 0.25 for both clusters.
The models in our analysis to follow, which only use the isochrones as a tracer of main sequence mass as a function of $M_G$, are almost invariant to the adopted age and metallicity
since we restrict ourselves to stars well below the MS turnoff. It is known \citep{Brandner2023} that the MESA isochrones provide a poor fit (ie. are under-luminous) for single stars with masses, 
$0.25 \lesssim M \lesssim 0.85 \, {\rm M}_\odot $, so we adjust the $M_G$ magnitude of the isochrones to match the MS ridge line. 

We do an initial coarse  filtering of obvious outlying data points blue-ward of the MS ridge line and red-ward of the equal-mass binary main sequence, and truncate using the binary mass-ratio track at the top and bottom of the main sequence.
To avoid confusion between binary stars and single stars leaving the main sequence, we make an upper-MS cut to our data at $M_G = 3$. We make the lower cut at $M_G = 9$, where the CMD main sequences are still very narrow and obvious.

\begin{figure}
\setlength\columnsep{3pt}
\begin{tabular}{cc}
    \includegraphics[width=0.48\columnwidth]{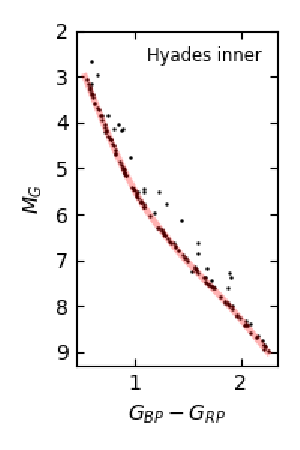}\hfill
    \includegraphics[width=0.48\columnwidth]{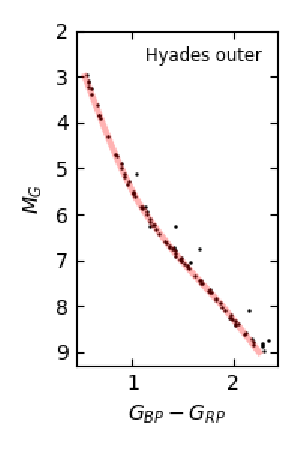} \\
    \includegraphics[width=0.48\columnwidth]{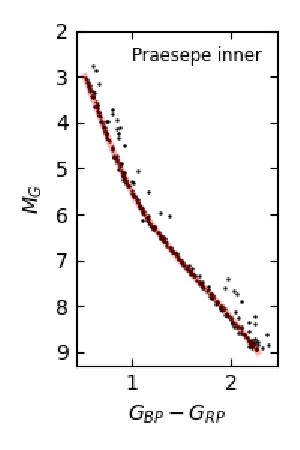}\hfill
    \includegraphics[width=0.48\columnwidth]{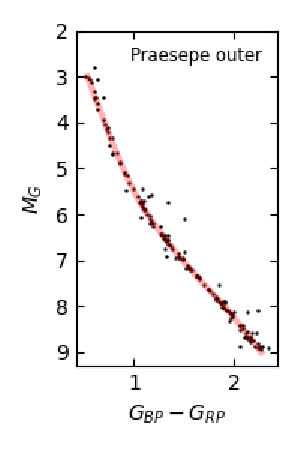}\\
 \end{tabular}
\caption{Colour-magnitude diagrams for the inner and outer regions of each cluster. The red lines are the magnitude-corrected MESA isochrones.}
\label{fig:CMD}
\end{figure}

To investigate possible segregation effects within each cluster, we divide each set of stars into samples located inside and outside of a sphere with radius equal to the tidal radius. For Hyades, this is estimated as $r_t = 10$~pc \citep{Perryman1998}, and for Praesepe we adopt $r_t = 10.8$~pc \citep{Roser2019}. At the median distances of the clusters, these project as 12~deg and 3.3~deg respectively.  
Colour-magnitude diagrams for the final selections of stars for each cluster and region are shown in Fig.~\ref{fig:CMD}.

\section{Model}

Our probablistic generative mixture model for the CMD is described in \citet{Albrow2022}. The data for each cluster
is the set ${\bm D} = \{{\bm D_k}\}$ for stars $k$, occupying locations  ${\bm D_k} = (G-G_{RP},M_{G})_k^T$ on the CMD with
uncertainty covariances $\mathbfss{S}_k$. We employ a scaling for each data covariance,
$\mathbfss{S}_k \rightarrow h^2 \mathbfss{S}_k$, where $h = h_0 + h_1 (M_{G ,k} \, - M_{G,0})$,
with $M_{G,0}$ set as the central magnitude of the data.

Briefly, the likelihood is described by
\begin{equation}
\label{eqn:like}
P({\bm D_k} | {\bm \theta}) = (1 - f_{\rm B} - f_{\rm O}) P_{\rm S}({\bm D_k} | {\bm \theta}) + f_{\rm B} P_{\rm B}({\bm D_k} | {\bm \theta}) + 
f_{\rm O} P_{\rm O}({\bm D_k} | {\bm \theta}),
\end{equation}
where $P_{\rm S}({\bm D} | {\bm \theta}), P_{\rm B}({\bm D} | {\bm \theta}), P_{\rm O}({\bm D} | {\bm \theta})$ are respectively the
likelihood functions for single stars, binaries and outliers (contaminants), $f_{\rm B}$ and $f_{\rm O}$ are the fractions of binary stars and outliers, and ${\bm \theta}$ is the vector of model parameters. The likelihood function for single stars depends on a parameterised mass function. For binary stars, the likelihood depends
additionally on a parameterised mass-ratio distribution function. The mass and mass-ratio distribution functions are each represented as linear combinations of Gaussian basis functions, which, in combination,  map onto bivariate Gaussians on the CMD. These are used to compute the likelihoods for each ${\bm D_k}$ given its (covariant) uncertainty, and ultimately the total likelihood for a given ${\bm \theta}$.

The model has undergone several modifications or extensions since its initial construction that we detail below.

\subsection{Mass function}

\begin{figure}
\setlength\columnsep{3pt}
\begin{tabular}{cc}
    \includegraphics[width=0.48\columnwidth]{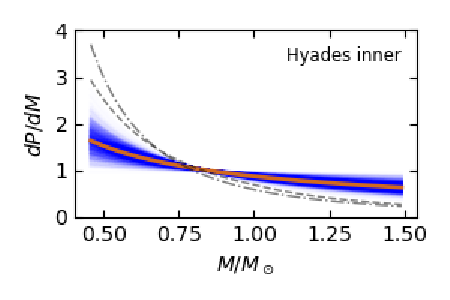}\hfill
    \includegraphics[width=0.48\columnwidth]{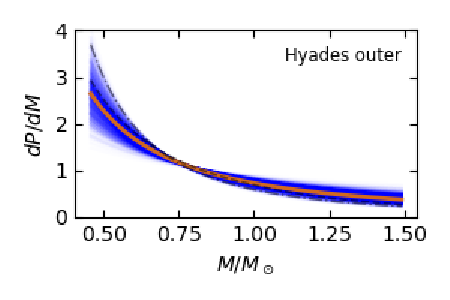} \\
    \includegraphics[width=0.48\columnwidth]{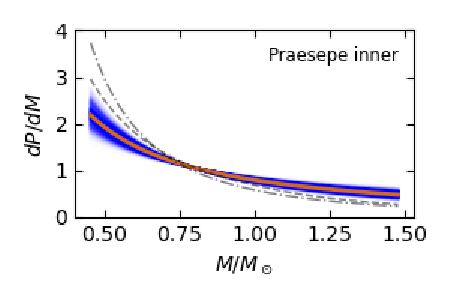}\hfill
    \includegraphics[width=0.48\columnwidth]{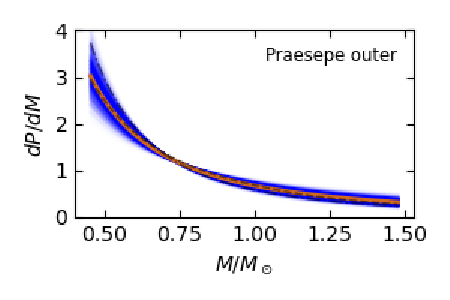}\\
 \end{tabular}
\caption{Mass function for the inner and outer regions of each cluster. The blue lines are random samples from the posterior distribution, and the red line indicates the posterior median probability model. Black lines are canonical mass functions from \citet{Salpeter1955} (dash-dot) and \citet{Chabrier2003} (dash).}
\label{fig:mf}
\end{figure}

The power-law form adopted for the mass function in \citet{Albrow2022} was found to be inadequate for some clusters. This may be because their stellar populations have undergone significant dynamical shaping since their formation, for instance from preferential ejection of lower-mass stars. We now adopt a more general form,
\begin{equation} \label{eqn:single_mass_function}
\begin{split}
\frac{dP(M|\gamma,k,M_{\rm 0},c_0,c_1)}{dM} =  
C_M \times \left( [(c_0 + c_1(M-M_{\rm min})] s_c  + M^{-\gamma} \right)  \times \\
\tanh(-k(M-M_0))  \times H(M - M_{\rm min}) \times H(M_{\rm max}-M),
\end{split}
\end{equation}
i.e. a power law plus a linear function in $M$ that is normalised by the scale factor $s_c = \left(\max(M_{\rm min},M_0)\right)^{-\gamma}$.
The final two terms truncate the function at (fixed) lower and upper bounds, $M_{\rm min}$
and $M_{\rm max}$,
and the $\tanh$ term allows the power law to roll-over at masses  $M \lesssim M_0$. A value of $M_0  \ll M_{\rm min}$ allows a
sharp cut to the power law with no roll-over.

\subsection{Mass-ratio distribution function}
\label{sect: MR-function}

In \citet{Albrow2022}, we adopted a polynomial form for the binary-star mass-ratio distribution function. We have explored using several alternative forms, including single and two-sided power laws, quadratics, and piecewise linear. We note that other papers, e.g.  \citet{Li2013, Li2020, Li2022}, have adopted strict power laws for the mass-ratio distribution.
Generally we have found that it is difficult to set non-informative priors on  these analytic parametric models, and that the forms of the
 analytic models impose undesirable constraints on the distribution. 
 
 We have finally adopted two different models for the distribution function. Our first model is linear sum of six shifted Legendre basis functions (five parameters). The inclusion of more components
 compared with \citet{Albrow2022} allows more flexibility, for instance the ability to curve upwards or downwards at either end of the distribution without affecting the other end. 
 
 Our second model is a histogram with $N_q$ equal-width bins for the distribution, i.e.
 \begin{equation}
\frac{dP(q | p_1, ..., p_{N_q})}{dq} = p_i, \quad q_i < q \leq q_{i+1}.
\end{equation}
There are $N_q$ parameters, $p_1 \, ... \, p_{N_q}$, for the model, with a normalisation constraint, $\Sigma_i p_i = N_q$.

\subsection{Binary fraction}

We have introduced a new parameter,  $\dot f_{\rm B}$, that allows the binary star fraction to vary with mass along the main sequence 
by $f_{\rm B} (M) = f_{{\rm B},0} + \dot f_{\rm B} (M_{\rm max} - M)$.

We emphasise that $f_{\rm B}$ is a consequence of parameters of the model, and should not be interpreted literally.
Since binaries with $q \lesssim 0.4$ barely deviate from the main sequence, such stars can contribute to either or both of the single or 
binary-star likelihood functions, depending on the adopted model basis. We regard $q = 0.5$ as a safe lower limit for $q$, beyond which 
binaries are well-distinguished from single stars in the Gaia photometric data sets in  this paper.

\subsection{Priors}

\begin{figure*}
\setlength\columnsep{3pt}
\begin{multicols}{5}
    \includegraphics[width=\linewidth]{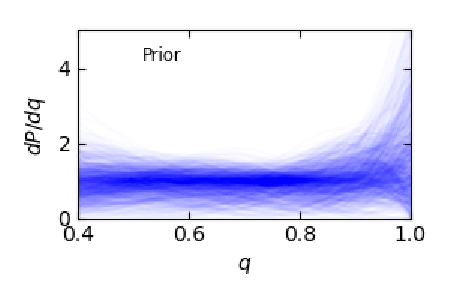}\par 
    \includegraphics[width=\linewidth]{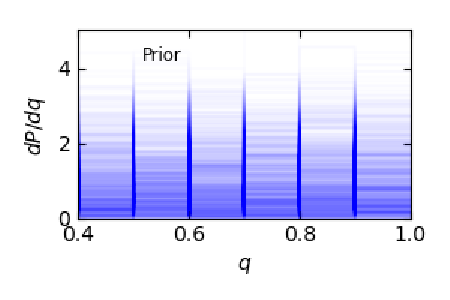}\\
    \includegraphics[width=\linewidth]{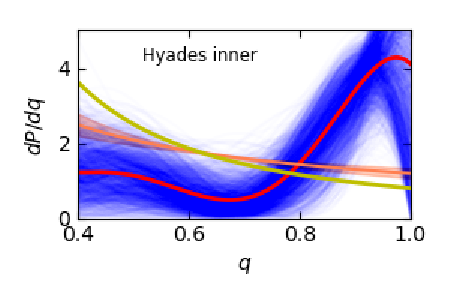}\par
    \includegraphics[width=\linewidth]{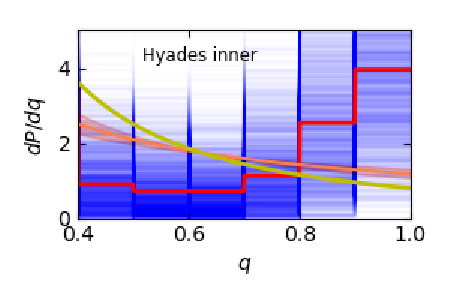}\\
    \includegraphics[width=\linewidth]{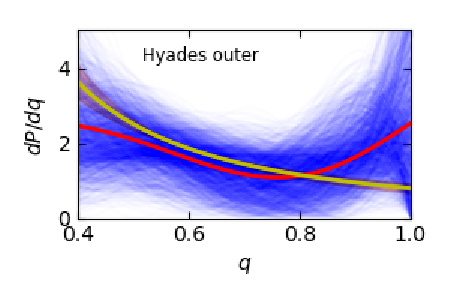}\par
    \includegraphics[width=\linewidth]{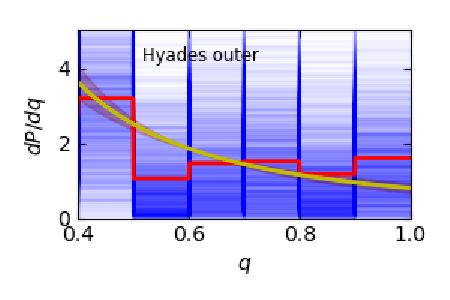}\\
    \includegraphics[width=\linewidth]{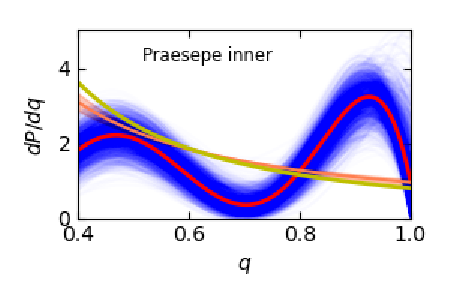}\par
    \includegraphics[width=\linewidth]{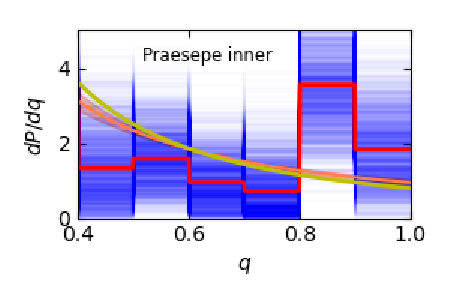}\\
    \includegraphics[width=\linewidth]{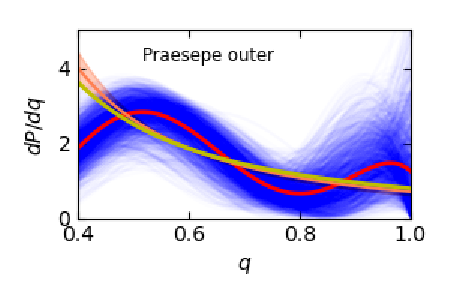}\par
    \includegraphics[width=\linewidth]{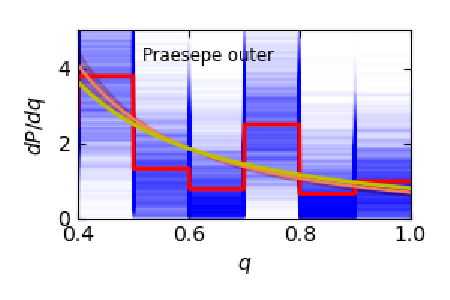}\\
    \end{multicols}
\caption{Mass-ratio distribution functions for the inner and outer binary-star populations of the two clusters (columns). The top row is the Legendre-function model, modified from \citet{Albrow2022}. The lower row shows the histogram models for $N_q = 10$ bins.  As in Fig.~\ref{fig:mf}, the blue lines are
random samples from the posterior distribution, and the red line is the {\em a-posteriori} maximum probability model.  Orange and yellow lines show respectively the implied $q$-distributions for 
randomly sampled stars from the measured mass function and the \citet{Chabrier2003} single-star galactic-disk mass function.
The first column shows the prior distribution.}
\label{fig:mqdist}
\end{figure*}

The complete model has up to ten  free parameters due to the mass function, binary and outlier fractions, and data error scaling, 
${\bm \theta} = (\log_{10} k, M_0, \gamma, c_0, c_1, f_{\rm B}, \dot{f_{\rm B}}, f_{\rm O}, \log h_0, h_1)$,
plus  either six or $N_q$  parameters for the mass-ratio distribution function.

Following Bayes theorem, the probability distribution for ${\bm \theta}$,

\begin{equation}
    P({\bm \theta} | {\bm D}) = \frac{P({\bm D} | {\bm \theta})  P({\bm \theta})}{Z},
\end{equation}

where $P({\bm D} | {\bm \theta})$ is the likelihood function, $P({\bm \theta})$ is the prior for ${\bm \theta}$ and $Z$ is the evidence (marginal likelihood), a constant for a given ${\bm D}$ and model.

To compute
the posterior probability distribution for ${\bm \theta}$, we must define a prior 
distribution, $P({\bm \theta})$. We have adopted sensible uniform priors for most parameters, however setting the priors for the  mass-ratio distribution function parameters
requires some thought.

As explained in \ref{sect: MR-function}, we now include six coefficients, $a_0 - a_5$ for the shifted-legendre polynomial model. The first of these, $a_0 = 1$, due to the normalisation condition. We draw each of the remaining coefficients, $a_1 - a_5$, from normal distributions, 
$a_i \sim \mathcal{N}(0, \sigma)$.  $\sigma$ is treated as a hyperparameter, which we draw from a reasonably-uninformative gamma distribution distribution, $\Gamma(2,3)$. 
This model strikes a natural balance between smaller values of $\sigma$ (which result in samples from $\mathcal{N}(0, \sigma)$ that have a higher prior probability) and higher values of $\sigma$ that may be favoured by the likelihood.
It thus chooses the smoothest and flattest functional forms that are consistent with the data,

For the histogram distribution, we would like to give equal prior weight to all histogram bins within the constraint, 
\begin{equation}
\Sigma_{i = 1}^{N_q} p_i = N_q , \quad  \quad \quad p_i \geq 0. 
\end{equation}
Naively  we might assign some prior permissible range to the first chosen
bin, $0 \leq p_1 < p_{\rm max}$, with the range for the second bin then being $0 \leq p_2 <   N_q - p_1$ etc. However, this procedure results in a selection-order bias, with a declining prior range for each subsequent bin. 

The correct equal-weighted prior for ${\bm p} = \{p_i\}$ is the Dirichlet distribution, ${\rm Dir}({\bm p})$ (see for instance \citet{Leja2017}). This is often described as the prior
for each sub-length of a fixed-length string cut into multiple pieces. 
If $N_q$ samples, $y_i$, are drawn from the Beta distribution,
\begin{equation}
    B(y) = \frac{\Gamma(N_q)}{\Gamma(N_q - 1)}  (1 - y)^{N_q-2} = N_q  (1 - y)^{N_q-2},
\end{equation}
then
\begin{equation}
p_i = \frac{y_i}{\Sigma_i y_i}
\end{equation}
are Dirichlet-distributed.

For our nested sampling procedure, we require the prior transform, which is the function that maps the range $[0,1]$ onto the prior. This function
is the inverse of the cumulative distribution, 
\begin{equation}
{\rm CDF}(p_i) = \int_0^{p_i} B(x) dx.
\end{equation}
This inverse  is also known as the quantile function or pixel percent function,
defined as $Q(p_i)$ such that ${\rm CDF}(Q) = p_i$.
By direct integration and normalisation, we find that
\begin{equation}
{\rm CDF}(p_i) =  1 - (1 - p_i)^{N_q-1}.
\end{equation}
After some algebra, this implies that
\begin{equation}
Q(p_i) = 1 - \left( 1 -  \frac{N_q-1}{N_q} p_i \right)^{1/(N_q-1)}.
\end{equation}

\section{Results}

We have used the dynamical nested sampler code DYNESTY \citep{Higson2019,Speagle2020} to sample the posterior parameter distribution for each cluster. The code has the ability to treat particular parameters, or combinations of parameters as constant, rather than free. As well as allowing all parameters to vary, we have made runs with combinations of $((\log k, M_0), c_0, c_1, f_{B,1}, \log_{10} h_0, h_1)$ frozen to constants $(\log k, M_0) = (4.1, 0.0)$ (an abrupt rather than $\tanh$ cutoff for the bottom of the mass function); $c_0 = 0$; $c_1 = 0$; $f_{B,1} = 0$; $\log_{10} h_0 = 0$; and  $h_1 = 0$.  We have also tested different values for $N_q$ in the range 4 to 10. Since all the different runs have different numbers of free parameters, we use the ratios of the evidence, $Z$, (Bayes factors) to decide which model or models are best. We follow the \citet{Kass1995} interpretation that $\Delta \log_{10} Z  > (0.5, 1, 2)$ represents (substantial, strong, decisive) evidence for a proposition. 

 We find in all cases that allowing scaled data uncertainties $\log_{10} h$ is necessary. Allowing 
$h$ to vary along the main sequence (i.e. using the $h_1$ parameter) is only necessary for the Hyades.
Allowing $f_B$ to vary along the main sequence is very mildly disfavoured in all cases.
A turnover in the mass function at the low-mass end (parameters $\log_{10} k$ and $M_{\rm 0}$)
was found to be unnecessary since we made sharp cuts at $M_G = 9$.
The evidence disfavoured using $c_0$ and $c_1$ for these data, so we also discard these parameters and use a pure power-law for the mass function.

In Figs.~\ref{fig:mf}  and~\ref{fig:mqdist} we show the mass-functions and mass-ratio distribution functions. 
For both clusters, the mass functions for the samples of stars outside of the tidal radius are consistent with the canonical mass-functions of 
\citet{Salpeter1955} and \citet{Chabrier2003} (which is very similar to \citet{Kroupa2001} in this mass range). Inside the tidal radii, the mass functions are much flatter. 

We show results for the mass-ratio distribution functions using the Legendre-function model and also as histograms with $N_q = 10$ bins. 
In order to be unaffected by different amounts of blending in the models between single and binary stars close to the main sequence, we renormalise 
these functions over $0.4 < q <= 1.0$. 
Our final models contain between 9 and 15 parameters, which is higher than the quantity of information we actually extract for 
interpretative purposes. Effectively we are marginalising (integrating) over some dimensions of the parameter space.
We choose to show the results based on these different model basis functions
to qualitatively illustrate the uncertainty in the distributions, and thus guard against over-interpretation of the fine details.

Assuming a power-law mass function, then for a given primary star, if the secondary star is randomly selected from the mass function, the mass-ratio distribution will follow the same 
functional form as the mass function. This form (implied by the particular mass function for each sample) is also shown in Fig.~\ref{fig:mqdist}. Additionally, we show the implied
$q$-distribution for randomly sampled stars from the \citet{Chabrier2003} single-star galactic-disk mass function, that we obtained through Monte Carlo sampling.
The samples of stars from the outer regions of each cluster are somewhat consistent with a randomly-chosen-secondary hypothesis. In contrast, the
samples of stars from within the tidal radius have a much higher frequency of high-mass-ratio binaries.

\begin{table*} 
        \centering
        \caption{Fraction of binary stars with mass ratio greater than a given $q$ for each cluster. The final rows, $FQ_{75}$ are the measured ratio of the number of binaries for which $q > 0.75$ to 
        the number for which $q > 0.5$, the single-star mass-function power law index, $\gamma$, and the value of $FQ_{75}$  implied by random sampling of the mass function. } 
        \label{table:f_tab_combined} 
	\begin{tabular}{|l|cccc|cccc|}
\hline
 & \multicolumn{4}{c}{Hyades}  & \multicolumn{4}{c}{Praesepe} \\
 & \multicolumn{2}{c}{$r < r_t$} & \multicolumn{2}{c}{$r \geq r_t$} & \multicolumn{2}{c}{$r < r_t$} & \multicolumn{2}{c}{$r \geq r_t$} \\
$q$ & legendre & histogram & legendre & histogram & legendre & histogram & legendre & histogram \\
\hline
0.5  & $ 0.125_{-0.028}^{+0.030} $ & $ 0.116_{-0.027}^{+0.029} $ & $ 0.08_{-0.03}^{+0.04} $ & $ 0.067_{-0.024}^{+0.030} $ & $ 0.128_{-0.016}^{+0.018} $ & $ 0.113_{-0.017}^{+0.020} $ & $ 0.119_{-0.022}^{+0.026} $ & $ 0.086_{-0.024}^{+0.026} $\\
\\
0.6  & $ 0.111_{-0.026}^{+0.029} $ & $ 0.105_{-0.024}^{+0.029} $ & $ 0.061_{-0.023}^{+0.030} $ & $ 0.055_{-0.021}^{+0.027} $ & $ 0.100_{-0.016}^{+0.018} $ & $ 0.093_{-0.016}^{+0.018} $ & $ 0.076_{-0.017}^{+0.022} $ & $ 0.067_{-0.020}^{+0.023} $\\
\\
0.7  & $ 0.104_{-0.025}^{+0.029} $ & $ 0.094_{-0.024}^{+0.027} $ & $ 0.046_{-0.019}^{+0.027} $ & $ 0.040_{-0.017}^{+0.023} $ & $ 0.090_{-0.017}^{+0.019} $ & $ 0.079_{-0.015}^{+0.017} $ & $ 0.047_{-0.016}^{+0.022} $ & $ 0.055_{-0.018}^{+0.022} $\\
\\
0.8  & $ 0.092_{-0.023}^{+0.026} $ & $ 0.078_{-0.022}^{+0.027} $ & $ 0.034_{-0.015}^{+0.024} $ & $ 0.025_{-0.012}^{+0.018} $ & $ 0.080_{-0.015}^{+0.016} $ & $ 0.068_{-0.014}^{+0.016} $ & $ 0.034_{-0.015}^{+0.021} $ & $ 0.022_{-0.010}^{+0.015} $\\
\\
0.9  & $ 0.057_{-0.015}^{+0.020} $ & $ 0.045_{-0.021}^{+0.025} $ & $ 0.020_{-0.010}^{+0.016} $ & $ 0.013_{-0.008}^{+0.014} $ & $ 0.044_{-0.008}^{+0.009} $ & $ 0.023_{-0.012}^{+0.014} $ & $ 0.021_{-0.009}^{+0.012} $ & $ 0.012_{-0.007}^{+0.011} $\\
\\
\hline
$FQ_{75}$ & $ 0.81_{-0.11}^{+0.07} $ & $ 0.76_{-0.11}^{+0.10} $ & $ 0.51_{-0.14}^{+0.16} $ & $ 0.52_{-0.16}^{+0.16} $ & $ 0.68_{-0.08}^{+0.08} $ & $ 
0.66_{-0.09}^{+0.09} $ & $ 0.34_{-0.12}^{+0.15} $ & $ 0.46_{-0.12}^{+0.12} $\\ \\
\hline
$\gamma$ & $ 0.80_{-0.25}^{+0.25} $ & $ 0.82_{-0.24}^{+0.22} $ & $ 1.65_{-0.30}^{+0.30} $ & $ 1.64_{-0.30}^{+0.30} $ & $ 1.28_{-0.17}^{+0.17} $ & $ 
1.30_{-0.17}^{+0.17} $ & $ 1.90_{-0.24}^{+0.25} $ & $ 1.94_{-0.23}^{+0.23} $\\ \\
Implied $FQ_{75}$ & $ 0.35_{-0.02}^{+0.02} $ & $ 0.35_{-0.02}^{+0.02} $ & $ 0.28_{-0.02}^{+0.02} $ & $ 0.28_{-0.02}^{+0.02} $ & $ 0.31_{-0.01}^{+0.01} $ & $ 0.31_{-0.01}^{+0.01} $ & $ 0.27_{-0.02}^{+0.02} $ & $ 0.26_{-0.02}^{+0.02} $\\
\hline
\end{tabular}

\end{table*}

We note that there is a certain amount of degeneracy for mass ratios $q > 0.75$, as all such stars tend to lie along the equal-mass binary main sequence (see fig. 3 in \citet{Albrow2022}). As a simple way of representing the tendency (or not) of the mass-ratio distributions to rise towards higher $q$ (that integrates over the degeneracy), we introduce a metric, $FQ_{75}$, that we define as the ratio of the number of cluster binaries for which
$q > 0.75$ to the number for which $q > 0.5$. Values of $FQ_{75}$ are listed in Table~\ref{table:f_tab_combined}.

\begin{figure}
    \includegraphics[width=\columnwidth]{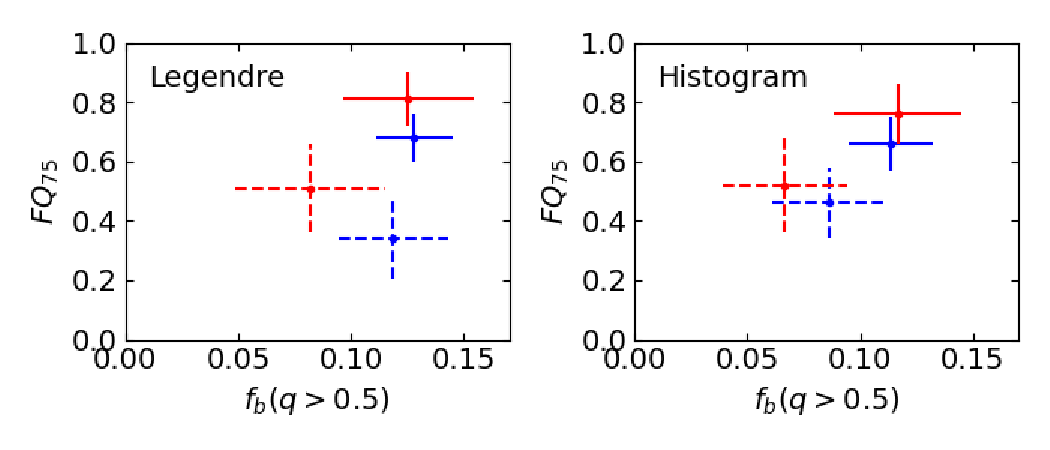}
\caption{Ratio of binaries with $q > 0.75$ to the number for which $q > 0.5$ plotted against overall frequency of binaries with $q > 0.5$ for Hyades (red) and Praesepe (blue).
Samples within the tidal radius are plotted with solid lines and those outside the tidal radius as dashed  lines. The left plot has been computed from the legendre model for the
mass-ratio distribution function, and the right plot has used the histogram model.
}
\label{fig:mr}
\end{figure}

In Fig.~\ref{fig:mr} we show  $FQ_{75}$ plotted against the overall fraction of observed cluster stars that are binaries with $q > 0.5$. There is a loose 
relationship, where the populations of stars with higher binary frequencies also have higher fractions of binaries that have $q > 0.75$. Referring to Table~\ref{table:f_tab_combined},
$FQ_{75}$ is higher for all populations than implied by random sampling of their present-day single-star mass functions, and in most cases higher than implied by 
random sampling of the Chabrier mass function (for which $FQ_{75} = 0.35$).

\begin{figure}
\setlength\columnsep{3pt}
\begin{tabular}{cc}
    \includegraphics[width=0.48\linewidth]{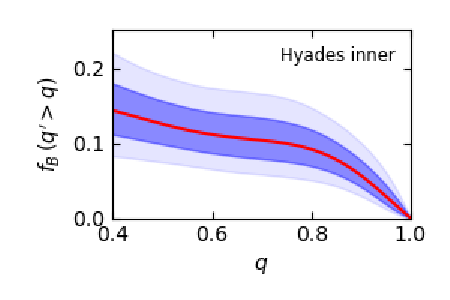}\hfill
    \includegraphics[width=0.48\linewidth]{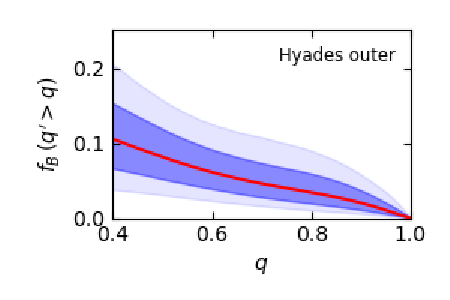}\\
    \includegraphics[width=0.48\linewidth]{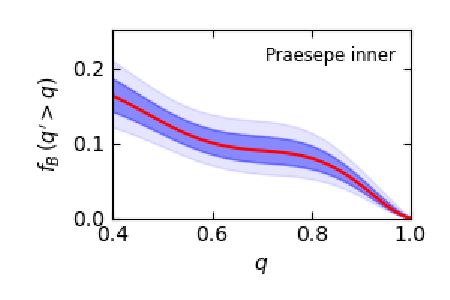}\hfill
    \includegraphics[width=0.48\linewidth]{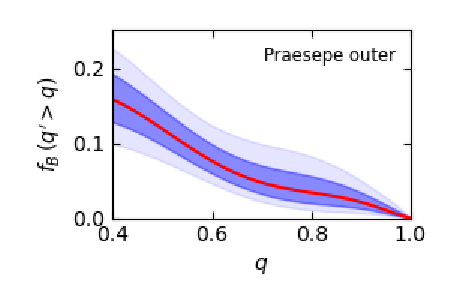}\\
\end{tabular}
\caption{Implied fraction of binary stars with mass ratio $q' > q$ for each cluster (red) together with its 1-$\sigma$ and 2-$\sigma$ uncertainties from the Legendre-function model.}
\label{fig:fbq}
\end{figure}

In Fig.~\ref{fig:fbq} we show the total fractions of observed cluster stars that are binaries with $q$ greater than a given value, based on the Legendre-function models. 
These fractions are obtained by integrating
the distributions in Fig.~\ref{fig:mqdist} backwards from $q = 1$. Numerical values of these functions for particular $q$ are given in Table~\ref{table:f_tab_combined} for both the Legendre and histogram bases.
The numbers are robust and and more-or-less independent of the choice of model basis and the other choices of including or excluding various parameters from the model. 

In Appendix~\ref{Appendix1}, Figs.~\ref{fig:realisation1} - \ref{fig:realisation4} each show seven random realisations of the CMD for each cluster.
These realisations have been generated by taking $(1-f_O) N$ random samples of the posterior maximum model, where $N$ is the number of stars in the
final observed CMD. They are shown next to the final observed CMDs for visual comparison. In all cases, the model CMD's give an excellent representation
of the observations. These simulations show that any apparent gaps in the observed distributions of binary stars in the CMD's are likely stochastic not physical.

\section{Conclusions}

From the observational results above we note the following:

\begin{enumerate}
\item
The inner parts of both clusters have mass functions that are flatter than a canonical power-law IMF. Stars  from outside the tidal radius that are evaporating from the clusters
have steeper mass functions, similar to field stars in the Galactic disk \citep{Chabrier2003}. 
For both inner and outer samples, the mass functions for Praesepe are a little steeper than
for Hyades.
These observations are consistent with a scenario that the clusters formed with mass functions flatter than that of  \citet{Chabrier2003} or \citet{Kroupa2001}.
Both clusters are undergoing segregated mass loss, with lower-mass stars being lost from the clusters at a greater rate then higher-mass stars, thus
steepening the mass function for ejected stars and flattening it for retained stars. 

\item
The overall  frequency of binaries with mass ratio $q > 0.5$ is 50 - 75 per cent higher within the tidal radius of Hyades than outside the tidal radius. 
For Praesepe, this frequency is 5 - 30 per cent higher within the tidal radius.

\item
There is a much higher relative frequency of higher-mass-ratio binaries within the tidal radius of each cluster.
The fact  that low-$q$ binaries are less-prevalent than high-$q$ binaries in the interior samples is perhaps a result of dynamical processing, 
with a higher rate of three-body interactions in the cluster interiors.
Such interactions can harden binaries or exchange binary companions, with both processes resulting in a kinetic energy transfer to recoil of the binary and/or single star.
The exchange process (sometimes via the intermediate phase of a triple system) is more-likely to retain the two highest-mass stars as a binary \citep{Heggie1975} thus converting lower-$q$ binaries to higher-$q$ . The ejected least massive star may leave the cluster core, accentuating the single-star mass segregation.

\end{enumerate}

%%%%%%%%%%%%%%%%%%%%%%%%%%%%%%%%%%%%%%%%%%%%%%%%%%

%\section*{Acknowledgements}
%
%We are grateful for ...

\section*{Data and Code Availability}

Gaia Data Release 3 (DR3) data, and the Gaia Catalogue of Nearby Stars are publicly available via the Gaia archive,  https://gea.esac.esa.int/archive/ ,
and the Centre de Donn\'ees astronomiques de Strasbourg(CDS) catalogue service, https://vizier.cds.unistra.fr/viz-bin/VizieR.

The CMDFITTER code used for this analysis is written in PYTHON and CUDA (via the pyCUDA python library). 
CUDA is an extension to C/C++ that uses an NVIDIA graphical processing unit to perform
parallel calculations. The code 
is available at \url{https://github.com/MichaelDAlbrow/CMDFitter}.



\bibliographystyle{mnras}
\bibliography{references} % if your bibtex file is called example.bib

%\begin{thebibliography}{99}
%\end{thebibliography}

%%%%%%%%%%%%%%%%%%%%%%%%%%%%%%%%%%%%%%%%%%%%%%%%%%

%%%%%%%%%%%%%%%%% APPENDICES %%%%%%%%%%%%%%%%%%%%%

\appendix

\section{Realisations of the the models}
\label{Appendix1}

\begin{figure*}
    \includegraphics{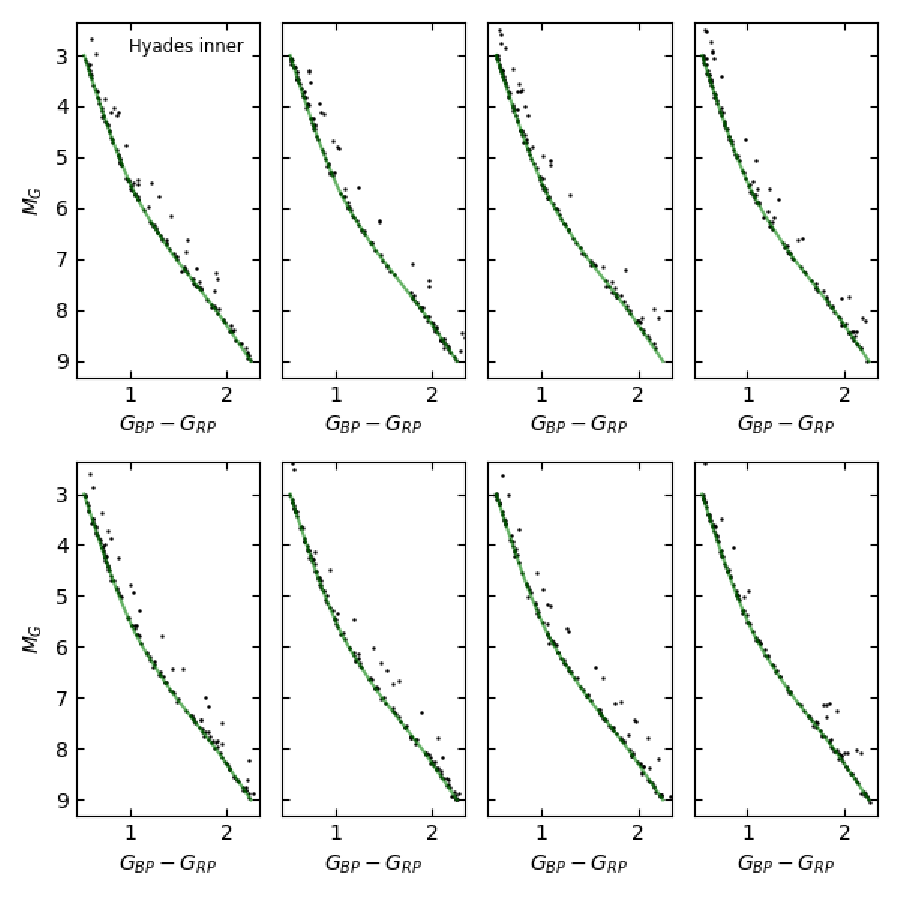}
\caption{The top left panel shows the final CMD for the Hyades inside a 10.0-parsec radius, with the colour-adjusted isochrone in green. The remaining seven panels show random realisations of the posterior maximum model.}
\label{fig:realisation1}
\end{figure*}

\begin{figure*}
    \includegraphics{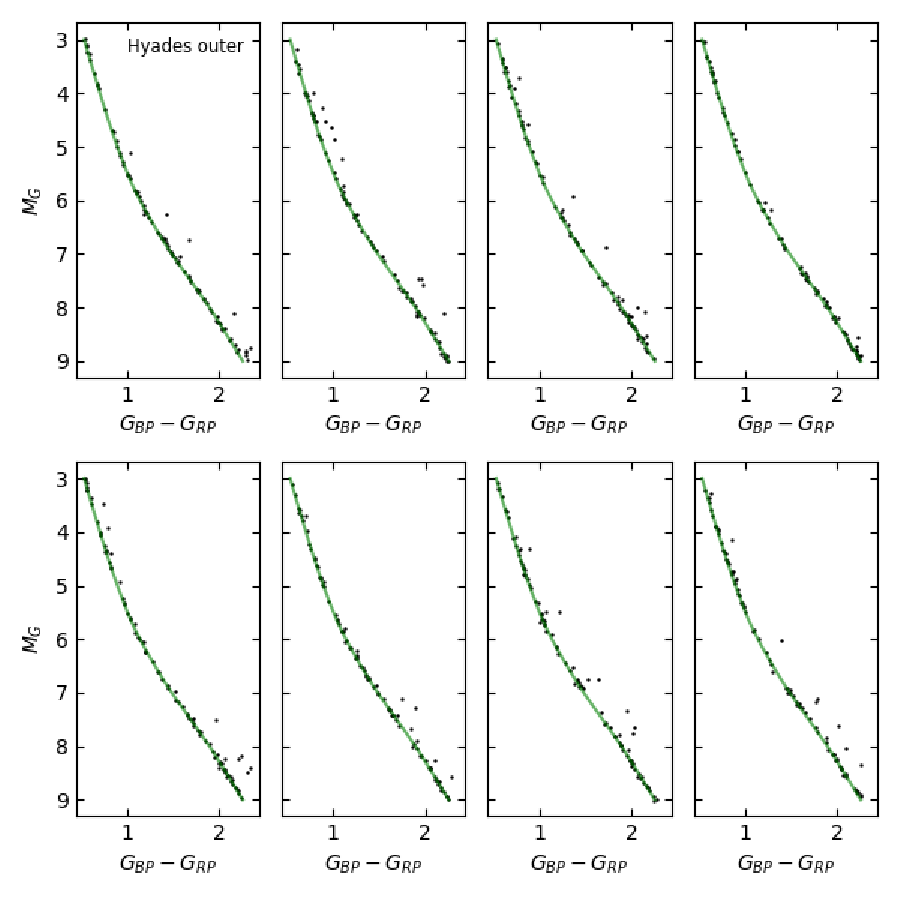}
\caption{The top left panel shows the final CMD for the Hyades outside a 10.0-parsec  radius with the colour-adjusted isochrone in green. The remaining seven panels show random realisations of the posterior maximum model.}
\label{fig:realisation2}
\end{figure*}

\begin{figure*}
    \includegraphics{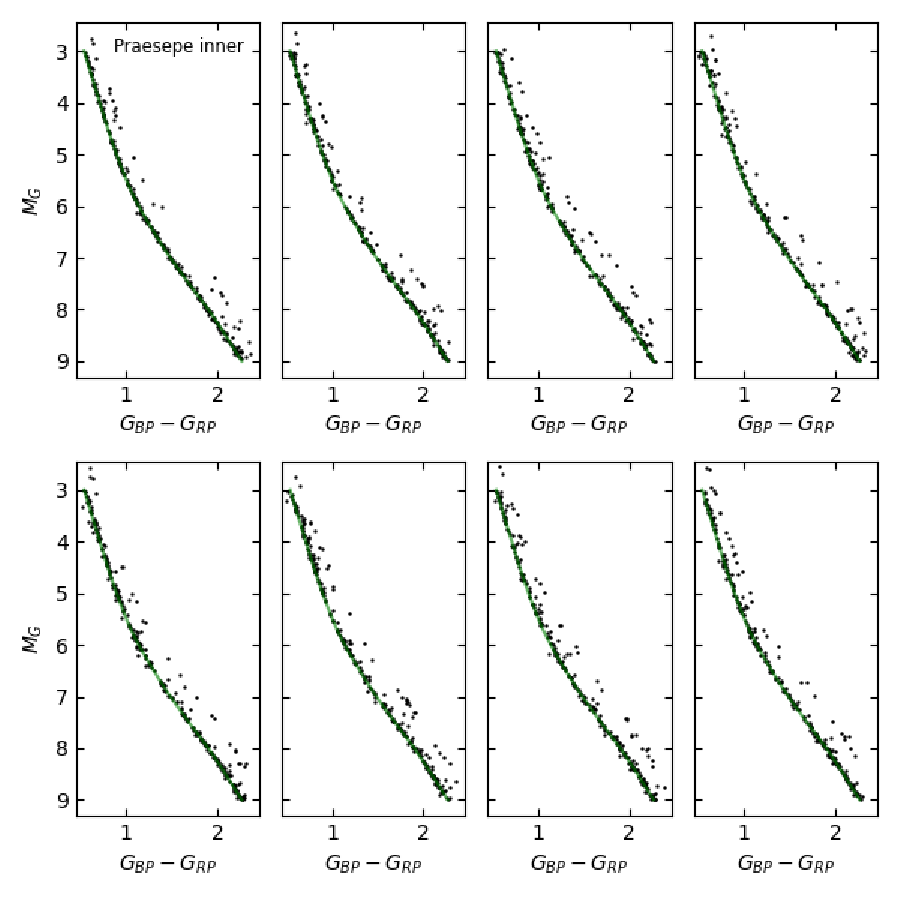}
\caption{The top left panel shows the final CMD for Praesepe inside a 10.8-parsec radius with the colour-adjusted isochrone in green. The remaining seven panels show random realisations of the posterior maximum model.}
\label{fig:realisation3}
\end{figure*}

\begin{figure*}
    \includegraphics{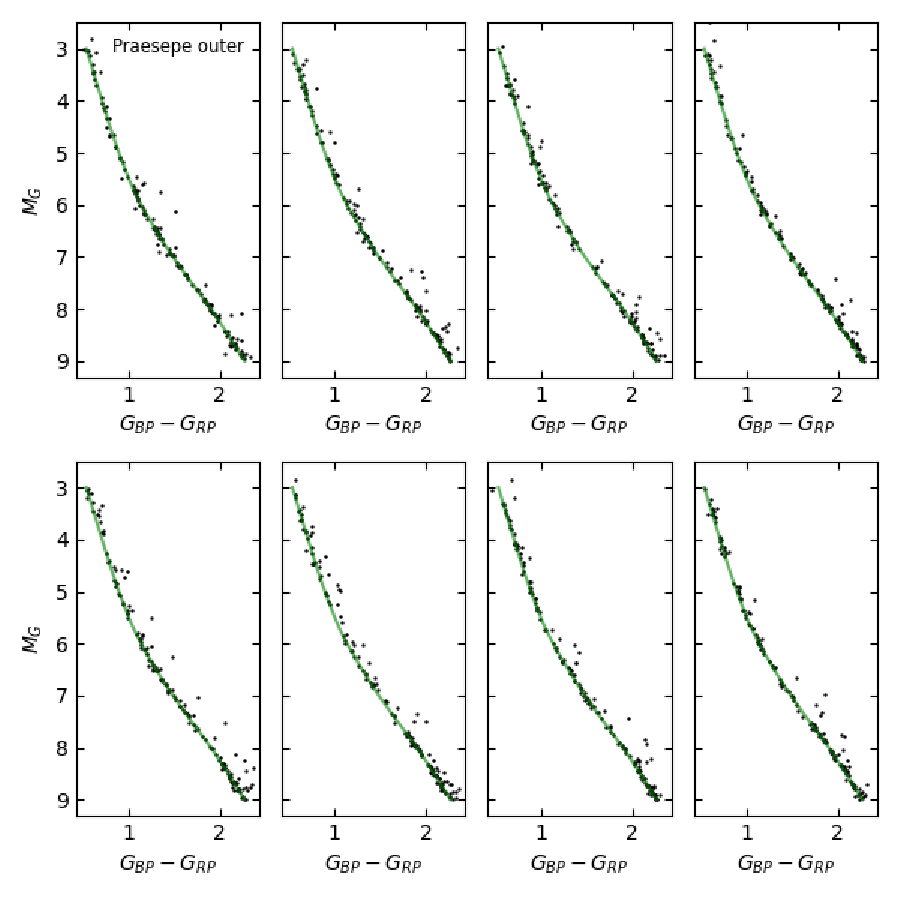}
\caption{The top left panel shows the final CMD for Praesepe outside  a 10.8-parsec radius with the colour-adjusted isochrone in green. The remaining seven panels show random realisations of the posterior maximum model.}
\label{fig:realisation4}
\end{figure*}

%%%%%%%%%%%%%%%%%%%%%%%%%%%%%%%%%%%%%%%%%%%%%%%%%%

% Don't change these lines
\bsp	% typesetting comment
\label{lastpage}
\end{document}